\documentclass[prl,aps]{revtex4}
\usepackage{epsfig}
\usepackage{graphicx}

\begin{document}

\title{Agterberg, Zheng, and Mukherjee Reply}

\maketitle

While Ikeda agrees that the crisscrossing lattice phase we have
suggested exists in principle, he argues that this phase does not
occur in CeCoIn$_5$, and he  further argues that our analysis is
lacking in a firm theoretical basis \cite{ike08}. In particular,
with respect to CeCoIn$_5$, Ikeda argues that the existing
experimental data provide stronger support for the LO vortex
lattice phase than the crisscrossing lattice phase. However, it
appears that recent experiments indicate that a typical FFLO phase
is unlikely in CeCoIn$_5$ \cite{ken08} (see below) so that neither
of the two vortex lattice phases can appear. With respect to the
theoretical basis, we address each of the issues raised by Ikeda
below and show that our theory does have a firm theoretical basis.
We now turn to a more detailed discussion of each of these points.

Recent experiments indicate that CeCoIn$_5$ does not have a usual
FFLO phase. In particular, neutron scattering experiments reveal
that spin density wave (SDW) order develops in the phase that was
previously thought to be a FFLO phase \cite{ken08}. The appearance
of this SDW order only within the superconducting state implies
the existence of a spatially modulated superconducting gap
function \cite{ken08}. The modulation of this gap function does
not change with field or temperature \cite{ken08}, in contrast to
what is expected for a usual FFLO phase. Furthermore,  the
spatially modulated gap function is argued to coexist with the
$d$-wave gap function \cite{ken08}, something that cannot occur
for a usual FFLO phase.

We now turn to the statement that our analysis is lacking in a
firm theoretical basis. We disagree with this statement. To reveal
why this is the case, we consider each of the three theoretical
arguments raised by Ikeda:

1- The statement of Ikeda: "the H$_{c2}$ transition in CeCoIn$_5$
at lower temperatures is discontinuous, implying that, according
to their criterion, the LO vortex phase is stable as the high
field phase" is incorrect. In particular, our criterion is derived
under the assumption that the transition is second order. A
different criterion arises when the transition is first order and
it can be shown that and the crisscrossing lattice phase can still
exist.

2-  Ikeda argues that our phase is characterized by two order
parameters $q$ and $\tau$, therefore there cannot be a second
order phase transition from the normal vortex phase into a
crisscrossing lattice phase. This statement is not obvious. In
particular, in the limit as $q$ becomes zero, it can be shown that
the parameter $\tau$ must also become zero (formally, in this
limit the fractional vortices that are responsible for the
crisscrossing vortex phase no longer exist). These two degrees of
freedom are strongly coupled in this limit. Only once $qL_z>>1$
(where $L_z$ is the length of the material) do these two
parameters become uncoupled. Since the limit that $q$ is small is
relevant to a transition from the normal vortex phase to the
crisscrossing lattice phase, one cannot preclude a single second
order transition.

3- The third issue raised by Ikeda relates to our assumption that
the gradient expansion perpendicular to ${\bf H}$ can be carried
out perturbatively. He argues that this can only be justified in
the limit of a large Maki parameter. He further argues that in
this limit, the lowest Landau level solution (which we have used)
is not permitted. These arguments are only correct for a very
specific set of assumptions. Namely, for a weak-coupling theory of
non-interacting electrons in the clean limit with a Fermi surface
that has a circular cross-section and a $\cos kz$ dispersion
perpendicular to this. We find that this set of criteria is highly
restrictive. In fact, the arguments of Houzet and Mineev show that
there exist weak-coupling theories for which our approach is
justified \cite{hou06}. Furthermore, there exist general arguments
that show that the crisscrossing lattice phase can exist
independent of the validity of the perturbation expansion.  In
particular, given that a superconductor develops a periodic
modulation, then there exist topological defects that combine half
a vortex with half a screw dislocation \cite{agt08}. The
crisscrossing lattice is a lattice of these defects.

In conclusion, we have shown that our crisscrossing lattice phase lies on a firm theoretical basis and have
pointed out recent experimental evidence that speaks against a FFLO phase in CeCoIn$_5$.\\

\noindent D. F. Agterberg, Z. Zheng, and S. Mukherjee\\
Department of Physics\\
University of Wisconsin-Milwaukee\\
Milwaukee, WI 53211

\end{document}